# STRANGELETS AT NON-ZERO TEMPERATURE

DAN MØNSTER JENSEN

and

JES MADSEN

*Institute of Physics and Astronomy, University of Aarhus*
*DK-8000 Århus C, Denmark*

ABSTRACT

A description of strange quark matter at finite temperature is given, with emphasis on the inclusion of finite size effects. For massless quarks all thermodynamic quantities can be evaluated analytically, while the case of massive quarks must be treated numerically. The resulting model is applied to the phase equilibrium between strange hadronic matter and strangelets, and some preliminary results are given.

## 1. Introduction

Low baryon number droplets of strange quark matter, so-called strangelets, might condense out of a hadronizing quark-gluon plasma in an ultra-relativistic heavy ion collision. The study of the ground state properties of strangelets is important in order to determine the possible (meta-) stability of these objects. However, even if strangelets are absolutely stable in the experimentally accessible region they may never be formed under the prevailing experimental conditions. Under which conditions we might expect to see strangelets cannot be inferred from a knowledge of the ground state properties of strangelets alone. In order to gain insight into the possible production of strangelets it is necessary to include the excited states through which the strangelet would pass at high temperature before ending up in its ground state. Using the MIT bag model[1] for the strangelet the strong interaction is mimicked by the bag constant and the bag boundary conditions. Further interactions can be included in perturbation theory, but we have set $\alpha_s = 0$ throughout. The single particle energy levels can be found by solving the Dirac equation with bag boundary conditions numerically. This is the shell model (see Madsen elsewhere in these proceeedings). An alternative approach is to compute a continous density of states based on the boundary conditions, which gives the liquid drop model. The latter approach is computationally much less expensive, and allows analytic evaluation of some quantities. Either of these models can be used to construct a partition function used for a statistical mechanical/thermodynamical description of strangelets. Shell model calculations have been carried out at finite temperature by Mardor and Svetitsky[2], but only in the case of vanishing baryon chemical potential. In the present paper only the liquid drop approach will be treated. A thorough study of the phase equilibrium between strange quark matter and strange hadronic matter was carried out by Lee and Heinz[3]. They also calculated isentropic expansion trajectories for a number of different initial conditions. This work is an attempt to extend their treatment to include the important finite size effects in the quark-gluon phase.

## 2. General Framework

In this section we describe the general framework for a statistical treatment of non-interacting quarks and gluons in an MIT bag, using the liquid drop model.

The starting point for the description is the grand canonical partition function, $Z$, for a collection of quarks, anti-quarks, and gluons obeying the confining boundary conditions of the MIT bag, which is considered to be *spherical*. In units where $\hbar = c = k_B = 1$, $Z$ is given by:

$$\log Z = \sum_i \sum_{\mathbf{k}_i} \xi_i \log(1 + \xi_i e^{-\beta(\sqrt{k_i^2 + m_i^2} - \mu_i)}) - \beta BV. \tag{1}$$

$B$ is the energy density of the perturbative vacuum inside the bag, known as the bag constant, and $\beta$ is the inverse temperature. The index $i$ runs over the species present in the bag. $\xi_i$ is $+1$ for fermions and $-1$ for bosons. The $\mathbf{k}_i$ sum runs over all single particle states. In the liquid drop model this sum is replaced by an integral, by introducing the density of states, $\rho_i(k)$, which as a consequence of the assumed spherical symmetry only depends on the length of the vector $\mathbf{k}$.

$$\log Z = \sum_i \xi_i \int_0^\infty dk\, \rho_i(k) \log(1 + \xi_i e^{-\beta(\sqrt{k^2 + m_i^2} - \mu_i)}) - \beta BV. \tag{2}$$

By applying a multiple reflection expansion[4,5] the density of states can be cast into the general form:

$$\rho_i(k) = g_i \left\{ \frac{1}{2\pi^2} k^2 V + f_S^{(i)}\left(\frac{m_i}{k}\right) kS + f_C^{(i)}\left(\frac{m_i}{k}\right) C + \cdots \right\}. \tag{3}$$

Here $V = \frac{4}{3}\pi R^3$ is the volume of the bag, $S = 4\pi R^2$ is the surface area, and $C = 8\pi R$ is the extrinsic curvature of the bag surface. The overall factor $g_i$ is the degeneracy related to the spin and colour degrees of freedom (6 for quarks and anti-quarks, and 16 for gluons). The volume term is seen to be universal, i.e. it depends neither on the particle species nor on the imposed boundary conditions. The contrary is true for the surface and curvature terms.

The surface term for quarks and anti-quarks has been known for some time[6] to be

$$f_S^{(q)}\left(\frac{m_q}{k}\right) = -\frac{1}{8\pi}\left\{1 - \frac{2}{\pi}\arctan\frac{k}{m_q}\right\}, \tag{4}$$

which vanishes in the limit $m_q = 0$. The surface term for gluons is zero. Thus for massless quarks, anti-quarks, and gluons the leading correction to the density of states is the curvature term.

The curvature term for massive quarks and anti-quarks, has not been derived using the multiple reflection expansion method, but recently Madsen[7] proposed the ansatz

$$f_C^{(q)}\left(\frac{m_q}{k}\right) = \frac{1}{12\pi^2}\left\{1 - \frac{3k}{2m_q}\left(\frac{\pi}{2} - \arctan\frac{k}{m_q}\right)\right\}, \qquad (5)$$

which has the correct limiting values for $m_q = 0$ and $m_q \to \infty$. For gluons the curvature term is[8]

$$f_C^{(g)} = -\frac{1}{6\pi^2}. \qquad (6)$$

All thermodynamic quantities are obtained through the (grand canonical) thermodynamic potential, $\Omega$, which, neglecting all higher order terms in Eq. (3) can be written

$$\begin{aligned}\Omega &= -T\log Z \\ &= \sum_i \{\Omega_{i,V}(T,\mu_i)V + \Omega_{i,S}(T,\mu_i)S + \Omega_{i,C}(T,\mu_i)C\} + BV.\end{aligned} \qquad (7)$$

In the case of zero temperature, the resulting expressions for $\Omega_{i,j}$ can be found in the contribution by Madsen to these proceedings.

For non-zero temperature we have not found an analytic expression valid in general. In the special case of zero quark mass, however, it is possible to evaluate the combinations $\Omega_{q,j} + \Omega_{\bar{q},j}$, for a single quark/anti-quark species, under the assumption $\mu_{\bar{q}} = -\mu_q$, which is valid when processes such as $q\bar{q} \leftrightarrow \gamma\gamma$ are in equilibrium.

$$\Omega_{q,V} + \Omega_{\bar{q},V} = -\frac{7\pi^2}{60}T^4 - \frac{1}{2}\mu_q^2 T^2 - \frac{1}{4\pi^2}\mu_q^4 \qquad (8)$$

$$\Omega_{q,C} + \Omega_{\bar{q},C} = \frac{1}{24}T^2 + \frac{1}{8\pi^2}\mu_q^2. \qquad (9)$$

The number densities are obtained as $n_{i,j} = -\left(\frac{\partial \Omega_{i,j}}{\partial \mu_i}\right)_T$, which for zero quark mass gives

$$n_{q,V} - n_{\bar{q},V} = \mu_q T^2 + \frac{1}{\pi^2}\mu_q^3 \qquad (10)$$

$$n_{q,C} - n_{\bar{q},C} = -\frac{1}{4\pi^2}\mu_q. \qquad (11)$$

The gluon contribution to the thermodynamic potential evaluates to:

$$\Omega_{g,V} = -\frac{8\pi^2}{45}T^4 \qquad (12)$$

$$\Omega_{g,C} = \frac{4}{9}T^2. \qquad (13)$$

The free energy is obtained as

$$F = \Omega - \sum_q N_q \mu_q, \qquad (14)$$

where $N_q$ is the number of quarks minus the number of anti-quarks of flavour $q$:

$$N_q = (n_{q,V} - n_{\bar{q},V})V + (n_{q,S} - n_{\bar{q},S})S + (n_{q,C} - n_{\bar{q},C})C. \qquad (15)$$

## 3. Quark Matter – Hadronic Matter Equilibrium

In this section we describe how to incorporate finite size effects into the treatment of phase equilibrium between quark matter and hadronic matter, at finite temperature.

### 3.1. Scenario

In the context of heavy ion collisions the formation of a strangelet will require a strangeness enriched plasma. If a quark-gluon plasma has a sufficient strangeness content it may cool to form a (meta-) stable strangelet.

Strange quarks are produced significantly only in strong interaction processes, i.e. as part of $s\bar{s}$-pairs. In order to develop a net strangeness the plasma must rid itself of some of the $\bar{s}$-quarks during the process of hadronization. This is thought to happen through $K^0, K^+$ emission[9]. We consider the resulting system, consisting of two phases, as having a fixed strangeness content, and as being in thermal, chemical, and mechanical equilibrium. This ideal situation is clearly not realized in heavy ion collisions, but serves as a reference for more sofisticated approaches as well as a testing ground for new ideas and models.

### 3.2. The Model

The quark-gluon phase is modelled as a collection of non-interacting $u, d, s$ quarks, their anti-quarks, and gluons in a spherical MIT bag.

A common $u, d$ mass, $m_q$, and chemical potential, $\mu_q$, is assumed. Collectively $u$ and $d$ will be referred to as the light quarks. There are thus only two independent chemical potentials, $\mu_q$ and $\mu_s$, or $\mu_B$ and $\mu_S$, where the latter are the chemical potentials for baryon number and strangeness respectively. The connections between these are

$$\mu_q = \frac{1}{3}\mu_B \qquad \mu_s = \frac{1}{3}\mu_B - \mu_S. \qquad (16)$$

The hadronic phase consists of the known hadronic spectrum, as obtained from the Particle Data Group[10] summary tables. Interactions are included by means of a hard core repulsion, leading to an excluded volume effect. The longer range attractive part of the potential has been neglected. There are several ways of including excluded volume effects[11], but we use the statistical bootstrap method of Hagedorn and Rafelski[12]. In this model thermodynamic quantities such as pressure, energy density, and baryon number density are all obtained from the corresponding ideal gas quantities by multiplication with the factor $(1 + \varepsilon^{(0)}/(4B))^{-1}$, where the superscript (0) denotes an ideal gas quantity, and $\varepsilon$ and $B$ are the energy density and the bag constant. Thus the pressure in the hadronic phase is given by

$$p^{(H)} = \frac{p^{(0)}}{1 + \frac{\varepsilon^{(0)}}{4B}}. \tag{17}$$

The corresponding point particle expressions are those of an ideal gas, given by

$$p^{(0)} = \sum_i p_i^{(0)}, \qquad p_i^{(0)} = \frac{g_i}{6\pi^2} \int_0^\infty dk \, \frac{k^4}{\varepsilon_i(k)} \frac{1}{e^{\beta(\varepsilon_i(k)-\mu_i)} \pm 1} \tag{18}$$

$$\varepsilon^{(0)} = \sum_i \varepsilon_i^{(0)}, \qquad \varepsilon_i^{(0)} = \frac{g_i}{2\pi^2} \int_0^\infty dk \, \frac{k^2 \varepsilon_i(k)}{e^{\beta(\varepsilon_i(k)-\mu_i)} \pm 1} \tag{19}$$

$$n_B^{(0)} = \sum_i b_i n_i^{(0)}, \qquad n_i^{(0)} = \frac{g_i}{2\pi^2} \int_0^\infty dk \, \frac{k^2}{e^{\beta(\varepsilon_i(k)-\mu_i)} \pm 1}, \tag{20}$$

where $i$ runs over the hadronic states. $g_i$ is the degeneracy factor of the species, and $b_i$ is the baryon number, which is $\pm 1$ for baryons/anti-baryons and zero for mesons. The signs in the denominator of the integrands correspond to Fermi-Dirac and Bose-Einstein statistics.

### 3.3. Equilibrium Criteria

The system consisting of a strangelet and a bulk hadronic phase is considered to be in chemical and thermal equilibrium, so the temperature and the strangeness- and baryon- chemical potentials are the same in the two phases

$$T^{(Q)} = T^{(H)}, \qquad \mu_B^{(Q)} = \mu_B^{(H)}, \qquad \mu_S^{(Q)} = \mu_S^{(H)}. \tag{21}$$

Minimizing the free energy leads to the mechanical equilibrium condition

$$\sum_i \left( \Omega_{i,V} + \frac{2}{R} \Omega_{i,S} + \frac{2}{R^2} \Omega_{i,C} \right) + B + p^{(H)} = 0. \tag{22}$$

This has the interpretation that the pressure exerted by the quarks and gluons in the bag is countered by the bag pressure, $B$, the hadronic pressure, and the surface tension and curvature force. Note that whereas the surface tension is always positive, and thus acts against the pressure, the curvature force can act both ways; depending on the value of the chemical potential. In equilibrium the chemical potential for a hadronic species, $i$, is given by its net light quark and strange quark content $\nu_{q,i}$ and $\nu_{s,i}$ as

$$\mu_i = \nu_{q,i} \mu_q + \nu_{s,i} \mu_s. \tag{23}$$

The equilibrium conditions Eq. (21) and Eq. (22) can then be solved numerically, yielding a set of triplets, $(T, \mu_q, \mu_s)$, defining the phase diagram.

## 4. Results

The quark phase is parametrized entirely in terms of the bag constant, $B$, and the quark masses. In almost all cases we have fixed the strange quark mass at $m_s = 150\,\text{MeV}$, while $u$ and $d$ have been assumed massless in all calculations. The bag constant has been varied between $B^{1/4} = 145\,\text{MeV}$, where $u, d$ matter is marginally stable in the bulk, and $B^{1/4} = 195\,\text{MeV}$, where $u, d, s$ matter is unstable relative to a gas of $\Lambda$'s, at zero temperature.

The hadronic phase is characterized in terms of the particle spectrum used, and the value of the bag constant. In most cases we have only included the ground state baryons $N, \Lambda, \Sigma, \Xi, \Omega$ and $\Delta(1232)$, while the mesons included are $\pi, K$ and $\eta$. This gives practically identical results compared to the case in which a full spectrum of all resonances below $\sim 2000$ MeV was used, except for $B^{1/4}$ well in excess of 200 MeV. All the results depicted here are for the limited spectrum.

The strangeness contents of the system is parametrized through the strangeness fraction, $f_s$, defined as the ratio of minus the strangeness density to the baryon density

$$f_s \equiv \frac{-n_S}{n_B}. \tag{24}$$

The results obtained for bulk systems are identical to those obtained by Lee and Heinz[3].

### 4.1. The Phase Diagram at $T = 0$

At zero temperature the quark thermodynamic potential is given by the analytic expressions in Madsen's contribution to these proceedings. The integrals occurring in the expressions for the hadronic properties can also be evaluated analytically. There are no non-strange mesons, and no anti-baryons. Solving the equilibrium conditions Eq. (21) and Eq. (22) gives the phase diagram in the $\mu_q$-$\mu_s$ plane as depicted in Fig. 1 for a *bulk* quark matter phase. The solid line is the coexistence line. Hadronic matter is the preferred phase inside the region bounded by the coexistence line, while the system is in the quark matter phase outside this region. Baryons are only present at $T = 0$ if the chemical potential, $\mu_i$, of the species is greater than the mass, $m_i$. This is the case in the region outside the dashed line. Each segment of the line is marked with a symbol denoting the particle species appearing when crossing the line from the inside. When the chemical potential for $\bar{K}^0, K^-$ becomes equal to the kaon mass, the kaons would undergo a Bose-Einstein condensation. Since this cannot easily be included in the model the region of the phase diagram to the left of the line marked $K$ is not accessible within the present framework. Kaons will be present to the right of this line only at finite temperature, so at zero temperature the region inside the dashed line will be devoid of hadrons. Hadrons may thus be present in the region between the full and dashed curves. Coexistence curves for lower values of $B$ will enclose a smaller region of the phase diagram. For values of $B^{1/4} \lesssim 165\,\mathrm{MeV}$ the coexistence curve will lie totally within the dashed curve, since quark matter will then be stable relative to a hadron gas. At the point where the phase coexistence line crosses the line where nucleons first appear there is a visible "kink" due to the fact that nucleons start to contribute to the pressure. The dashed-dotted line represents systems with a total strangeness fraction of 0.1. In the hadronic phase such a system would consist entirely of $N$'s and $\Lambda$'s, for the parameter values chosen.

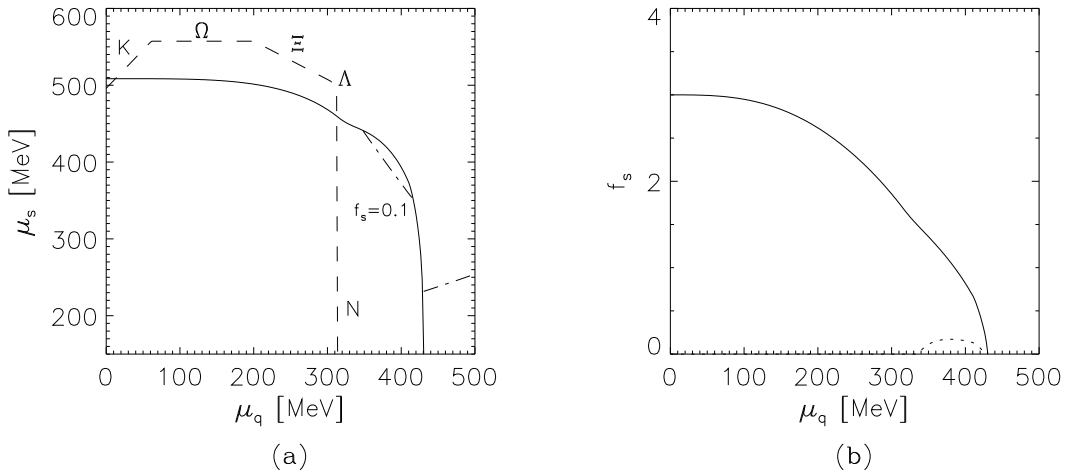

Figure 1: (a) Phase diagram for bulk strange quark matter and hadronic matter at $T = 0$. (b) The strangeness fraction in the quark (full line) and hadronic (dotted line) phases as a function of the light quark chemical potential. $B = (190\,\mathrm{MeV})^4$, $m_s = 150\,\mathrm{MeV}$. See text for further details.

Fig. 1 (b) depicts the strangeness fraction in the two phases along the coexistence curve, parametrized by the light quark chemical potential $\mu_q$. For a sufficiently high strangeness fraction there is no hadronic phase – with this choice of parameters – at zero temperature.

### 4.2. The Phase Diagram for $T \neq 0$

At finite temperature the quark matter phase contains thermal gluons and anti-quarks, while the hadronic phase contains non-strange mesons and anti-baryons. A phase diagram similar to the one in Fig. 1 can be constructed for a fixed temperature. The effect of increasing the temperature is that the hadronic phase occupies a smaller region of the phase diagram. This is because there is a large number of massless degrees of freedom in the quark matter phase, whereas the lightest constituent in the hadronic phase is the pion with a mass of $\sim 140\,\mathrm{MeV}$. This means that the pressure increases faster in the quark phase than in the hadronic phase, favouring the former over the latter. From this argument it is also clear that an increase in the bag constant will have the opposite effect, since $B$ effectively contributes to the hadronic pressure. The quark matter phase benefits from an addition of strange quarks, as opposed to the hadronic phase where the presence of strange hadrons tends to increase the free energy, because of their larger masses. Thus increasing the strangeness fraction will favour the quark phase. This effect is clearly seen in Fig. 2, where the phase diagram for a fixed strangeness fraction has been projected onto the $n_B$-$T$ plane. The solid curve is for a strangeness fraction of 0.1, whereas the dashed curve is for $f_s = 0$. The area between two like curves corresponds to a mixed phase. The quark-mixed phase boundary is the curve at high baryon density, and the other curve is the boundary between the mixed phase and the hadronic phase.

The effect of finite size of the quark phase is seen in Fig. 3, where phase diagrams for a bulk quark phase and a quark phase with baryon number $A = 20$ have been superimposed. The most notable difference is that the hadronic phase grows on account of the mixed phase when going from a bulk to a finite quark phase.

## 5. Discussion

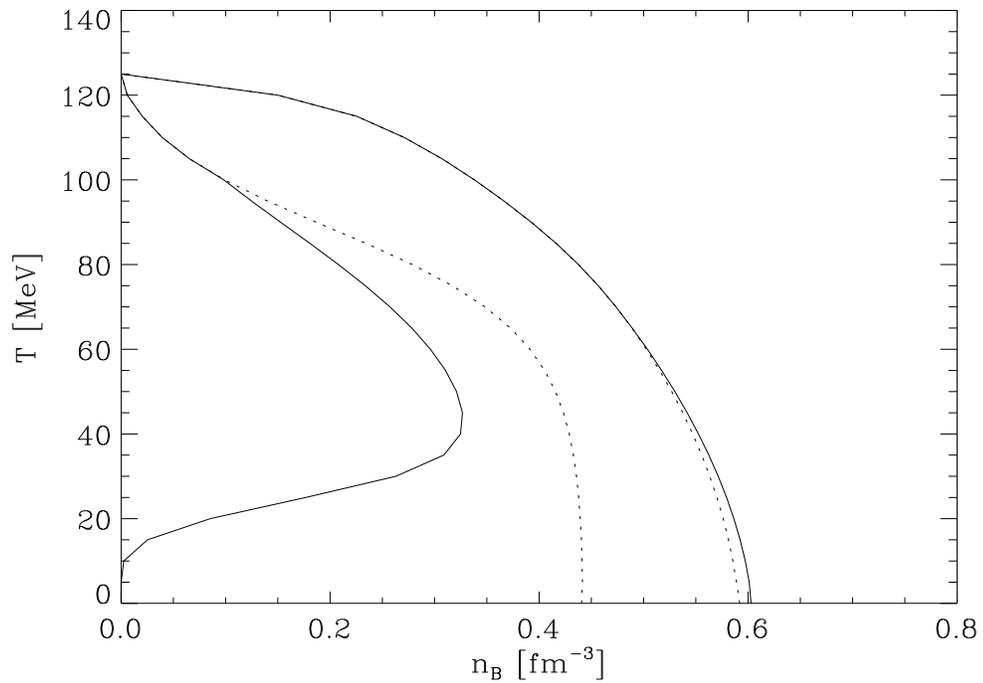

Figure 2: Phase diagram for a strangelet with baryon number $A = 20$, and a bulk hadronic phase. The full curve is for a strangeness fraction, $f_s = 0.1$, while the dashed curve is for $f_s = 0.0$ $B^{1/4} = 180\,\text{MeV}$, $m_q = 0$, $m_s = 150\,\text{MeV}$.

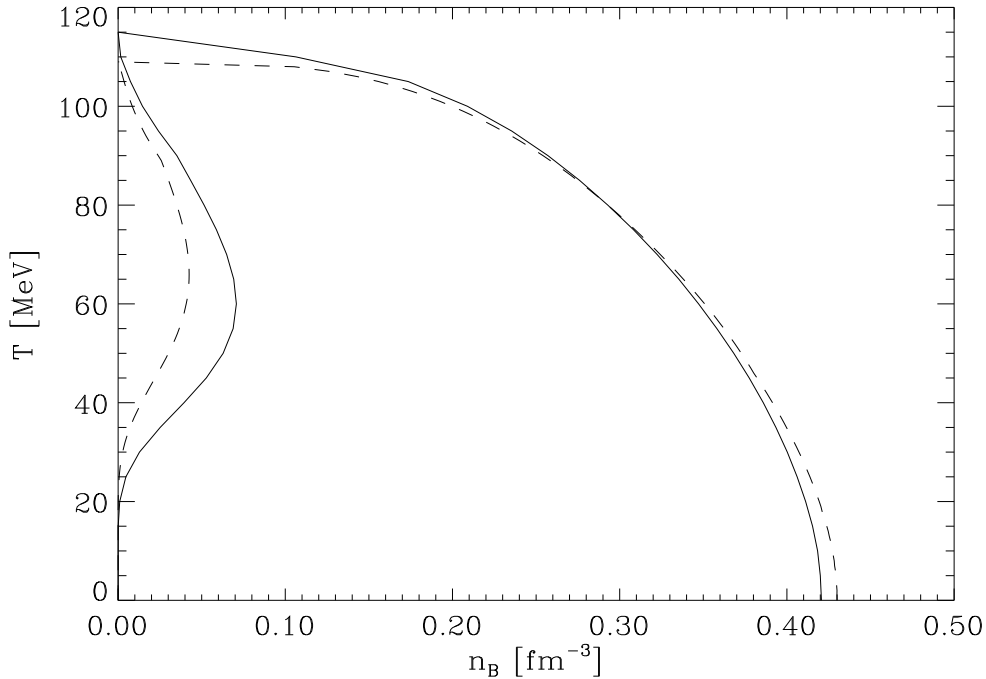

Figure 3: Phase diagram for a bulk hadronic phase in equilibrium with a bulk quark phase (dashed line), and a quark phase with baryon number $A = 20$ (full line). $B^{1/4} = 165\,\mathrm{MeV}$, $m_q = 0$, $m_s = 150\,\mathrm{MeV}$, $f_s = 0.1$.

We have described how to incorporate finite size effects into calculations on the equilibrium between strange quark matter and strange hadronic matter. The results displayed in Fig. 3 show a clear difference from the bulk case, but the detailed effects of this difference as a function of baryon number and other parameters remain to be investigated. The goal is to use such calculations to learn whether strangelets may be produced in heavy ion collisions.

The work described here is still in progress. The model presented can be improved upon in several aspects. As an example the partition function given in Eq. (1) counts all states, including those that are not colour singlets. A partition function describing only colour singlet states can be constructed using the group theoretical projection technique developed by Redlich and Turko[13].

## 6. Acknowledgement

This work was supported in part by the Theoretical Astrophysics Center under the Danish National Research Foundation.